\documentclass[10pt,letterpaper]{article}
\usepackage[top=1in,left=1in,right=1in,footskip=2.75in,marginparwidth=1in]{geometry}

\usepackage{csquotes}
\usepackage{gensymb}
\usepackage{enumitem}
\usepackage{nicefrac}
\usepackage{lastpage,fancyhdr,graphicx}
\usepackage[T1]{fontenc}
\usepackage{tgheros}
\usepackage{orcidlink}
\usepackage{amsmath}
\usepackage{amssymb}
\usepackage{booktabs}
\usepackage{comment}

\begin{document}

\begin{flushleft}
{\Large
\textbf\newline{Spatial DJ: Spatial Audio Control for Arbitrary Loudspeaker Layouts}
}
\newline
\\
Anastasia Geleta\textsuperscript{1,$\dagger$,*}, Margarita Geleta\orcidlink{0000-0001-5823-9776}\textsuperscript{1,2,*}
\author{Anastasia Geleta}
\\
\bigskip
$^1$\footnotesize{AVINCE, San Francisco, CA 94115, USA}\\
$^2$\footnotesize{Department of Electrical Engineering and Computer Science, University of California, Berkeley, Berkeley, CA 94720, USA}\\
$\dagger$ Corresponding author: \texttt{geleta@avince.es} \\
$^*$\footnotesize{A. Geleta and M. Geleta contributed equally to this work.}
\end{flushleft}

\begin{figure*}[ht!]
  \includegraphics[width=\textwidth]{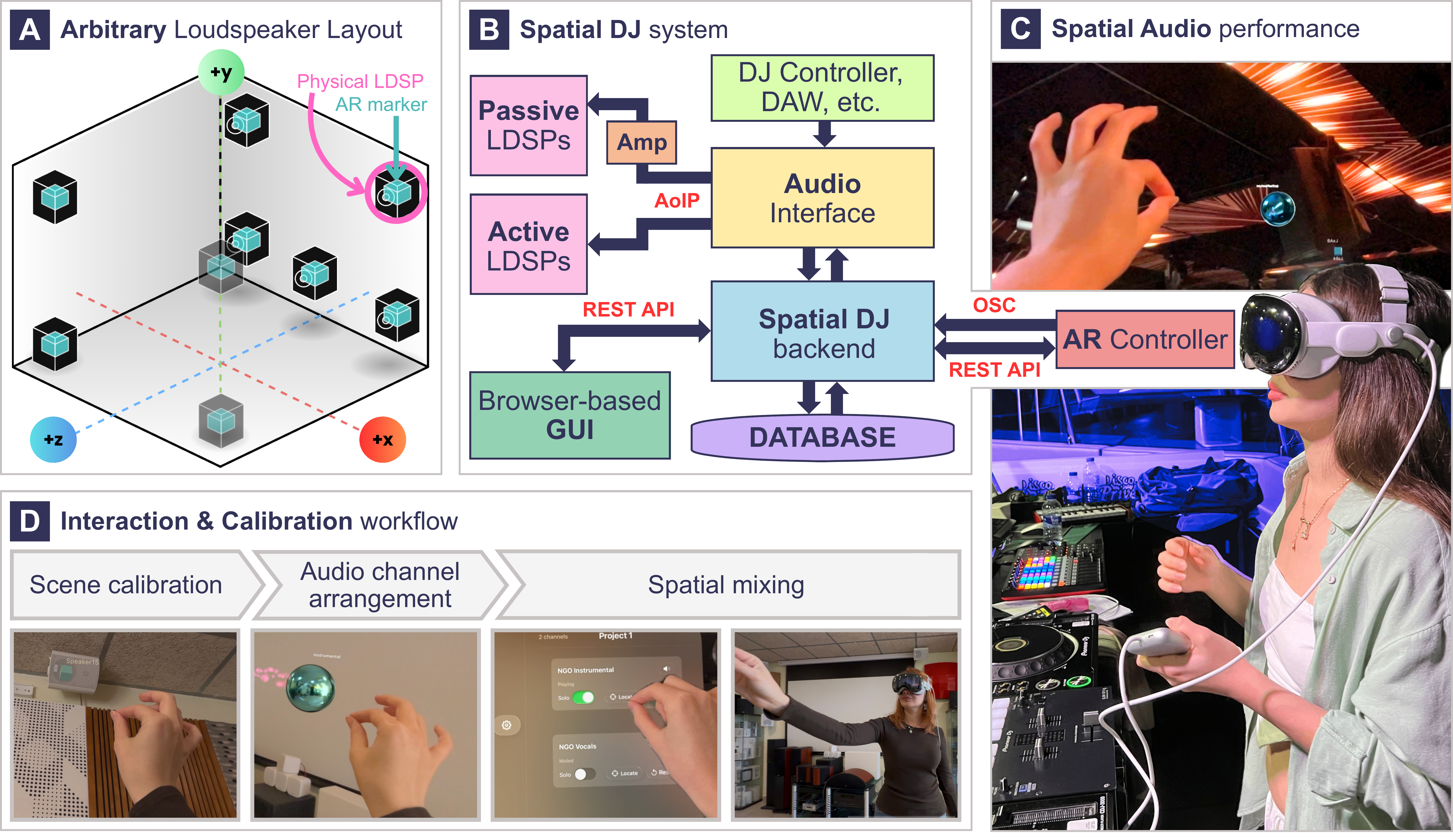}
  \caption{\textbf{\emph{Spatial DJ}}. \textbf{A}  Arbitrary loudspeaker layouts are registered in a room-aligned coordinate frame by aligning AR markers with physical loudspeakers. \textbf{B} The \emph{Spatial DJ} backend connects the AR controller, database, audio interface, and active or passive loudspeakers, while supporting external sources such as DJ controllers and DAWs. \textbf{C}  The AR interface enables creators to spatially position and manipulate virtual audio objects during live performance. \textbf{D} Interaction workflow from scene calibration and channel arrangement to spatial mixing.}
  \label{teaser}
\end{figure*}

\begin{abstract}
Object-based spatial audio promises \emph{render-anywhere} playback, yet deploying it in a new venue still requires two tasks that are rarely addressed together: acquiring the physical loudspeaker geometry and enabling creators manipulate spatial audio while listening through the target loudspeaker system rather than a headphone proxy. We present \emph{Spatial DJ}, a system for on-site loudspeaker layout acquisition and real-time spatial audio rendering and control, using head-worn AR in a coordinate frame aligned to the physical loudspeakers. In an 18-loudspeaker room, we evaluate calibration accuracy, repeatability, and control latency, and conduct an 18-participant listener study comparing diffuse and orbiting point-source rendering configurations applied to vocal and instrumental layers. Orbiting point-source conditions received higher ratings of perceived immersion, energy, and novelty than the diffuse baseline. Beyond this controlled setting, four representative deployments illustrate reuse of the same system across heterogeneous loudspeaker venues.
\end{abstract}


\section{Introduction}

Spatial audio has become an established production and interaction medium across music, film post-production, video games, virtual reality (VR), 360$\degree$ video, interactive sound art, and live production mixing \cite{murphy2010spatial, sinclair2020principles, stampfl2003ardj, 10.1007/s10055-015-0274-4, lyon2008spatial}. Studio workflows typically center on Digital Audio Workstations (DAWs) with panner plug-ins and export paths to channel-based and object-based formats \cite{robinson_2025_dolby, dolby_panner}. Venue-scale systems often rely on specialized renderers such as SPAT by FLUX:: SE and Ircam Amplify \cite{Flux2025SPATRevolution}, while gaming and VR pipelines bind object metadata to in-scene transforms through game engines (e.g., Unity 3D \cite{UnityTechnologies_UnityEngine_2026} or Unreal Engine \cite{EpicGames_UnrealEngine_2026}) and middleware (e.g., Apple's Physical Audio Spatialization Engine \cite{US11172320B1_phase}). Despite these increasingly capable toolchains, a workflow mismatch can arise when  spatial soundscapes are authored on a screen \cite{mathew2017survey, Engeln2025SpatialAudioPaths} or monitored via binaural previews \cite{audiomixr_2025_geleta}, but are ultimately intended to be reproduced in a physical venue with its own loudspeaker geometry, room acoustics, routing constraints, and audience position. 
 
This mismatch can become a deployment challenge when the target loudspeaker geometry is not fixed. Many production and reproduction approaches implicitly assume known, symmetric layouts or an optimal listening position, while in real venues loudspeaker layouts vary, equipment may be imperfectly documented, and setup time is limited \cite{Peters2011CurrentTechnologies}. At the same time, \emph{render-anywhere} paradigms in object-based spatial audio increasingly decouple the spatial sound scene description from the playback system \cite{mpegh2015, rumsey2017surround}: a renderer can, in principle, adapt object-based content to a range of loudspeaker configurations. In practice, however, deploying object-based content in a new venue still involves two aspects that are rarely addressed together: first, obtaining the loudspeaker geometry on site, since many renderers assume measured or configured loudspeaker coordinates \cite{Flux2025SPATRevolution}, and second, providing a design and control interface through which creators can iterate while standing in the room and hearing the actual loudspeakers rather than a headphone proxy.

This gap is particularly relevant outside controlled studio environments. Creators may need to bridge multiple coordinate systems (software scene, room coordinates, loudspeaker placement) and multiple control layers (authoring User Interface (UI), renderer, routing), potentially involving ad hoc calibration and manual iteration. Working spatial audio venues, such as \emph{Audium SF} \cite{Audium2026Residency} or \emph{Envelop} \cite{EnvelopArtistResidency}, usually require creators to use custom audio production stacks through their residency programs. Recent eXtended Reality (XR) interfaces show the promise of spatial authoring \cite{Deacon2023SpatialDesign}: systems such as \emph{AudioMiXR} support direct six degrees of freedom (6DoF) manipulation of virtual audio objects in the surrounding environment, but target head-tracked binaural playback for a single user \cite{audiomixr_2025_geleta}. Other approaches couple spatial interaction to physical loudspeakers: \emph{ARCube} uses an AR-based 3D panning workspace within a fixed 28-loudspeaker installation \cite{arcube2024}, while \emph{boomRoom} demonstrates embodied manipulation of virtual audio sources using Wave Field Synthesis (WFS) \cite{the_boomroom_2014}. What remains less explored is how such interaction can be coupled to loudspeaker layouts that vary across deployment sites. This leaves not only an operational question of how such interfaces can be coupled to changing physical loudspeaker layouts, but also a perceptual question: when different spatial renderings of the audio content are reproduced over a physical loudspeaker environment, how do listeners experience them? Characterizing these differences can inform the design of spatial controls in future 3D audio authoring interfaces.

We present \emph{Spatial DJ}---a spatial audio rendering and control system designed for the messy realities of spatial audio deployment (Fig. \ref{teaser}). It combines a backend renderer, multichannel audio routing, and a frontend composed of a desktop and Augmented Reality (AR) control interfaces. The system co-locates interaction and audition by registering physical loudspeaker positions to a room-aligned coordinate frame, allowing users to manipulate virtual audio objects while the corresponding audio is rendered through the physical loudspeaker array in real time. To enable this on-site layout acquisition, an AR calibration workflow is used to align virtual speaker markers to physical loudspeakers and persist the registered geometry for reuse. In our implementation, the \emph{Spatial DJ} AR controller runs on Apple Vision Pro. Given the calibrated loudspeaker coordinates, the backend maps lightweight object state to per-speaker gains weighted by distance with constant power normalization, supporting point-source audio objects which are spatialized by distance, as well as ambient objects distributed equally across all loudspeakers. Rather than assuming a fixed loudspeaker geometry, \emph{Spatial DJ} applies the same object representation, spatial audio controls, and gain mapping rules to user-registered loudspeaker layouts. 
\newpage
We investigate two research questions (RQs):
\begin{enumerate}[label=\textbf{RQ$_\arabic*$:}, leftmargin=*]
    \item Can \emph{Spatial DJ} support repeatable on-site layout acquisition and responsive spatial audio rendering across changing loudspeaker geometries?
    \item When rendered and authored with \emph{Spatial DJ}, how do different spatial rendering configurations applied to different audio objects shape listeners' perception?
\end{enumerate}

We evaluate the system with a focus on engineering feasibility, deployment practicality, and listener perception. In a controlled 18-loudspeaker room, we quantify loudspeaker calibration accuracy and repeatability across operators and measure control latency under varying runtime conditions. We then use the same room to study the perceptual consequences of modifying spatial parameters (position and spread) of two different audio objects (vocal and instrumental) by comparing four predefined spatial rendering configurations: a diffuse baseline, vocal orbit, instrumental orbit, and dual orbit conditions. In a listener study with 18 participants, we compare ratings of perceived immersion, motion clarity, vocal focus, energy, novelty, and distraction or discomfort across these configurations. Finally, we complement these controlled evaluations with four representative deployments across distinct loudspeaker layouts, documenting how the same calibration, routing, control, and rendering were applied across heterogeneous venues.

\section{Related Work}

\subsection{Spatial Audio Rendering over Physical Loudspeakers}

Traditional spatial audio systems often assume idealized reproduction conditions such as fixed, symmetric loudspeaker layouts and a designated optimal listening position \cite{Flux2025SPATRevolution}, known as the \emph{sweet spot}, which is approximately an equidistant point from all loudspeakers \cite{Kendall2010Spatial}. Vector-Base Amplitude Panning (VBAP), for instance, computes gain ratios for the nearest loudspeaker pair or triad given a virtual audio object position and a precise geometric model of the speaker setup \cite{pulkki1997virtual, mathew2017survey}. This procedure yields localized phantom imagery as listeners tend to perceptually construct continuity in moving sounds even when the underlying spatial rendering itself is discrete \cite{Barrett2012HighOrderAmbisonics}. Even so, across venue scales, loudspeaker placement and coverage remain important for avoiding perceptual gaps \cite{Harrison1998SoundSpaceSculpture}. WFS can produce stable localization over an extended listening area by recreating physical wavefronts \cite{the_boomroom_2014, soundholo_2025}, but it typically relies on dense, carefully arranged loudspeaker arrays and controlled reproduction conditions, which constrains its practical deployment outside bespoke installations. 

To relax these constraints, alternative rendering techniques have been developed which accommodate irregular or ad hoc loudspeaker layouts. Distance-Based Amplitude Panning (DBAP) does not require a symmetric or regularly spaced loudspeaker layout or specified listener location. Instead, it weights loudspeakers as a function of distance to the virtual audio source, without requiring an equidistant ring or a single sweet spot \cite{lossius2009dbap}. Typical formulations assume that every speaker contributes regardless of source position, making DBAP attractive for concerts, installations, and other contexts where there is no predefined symmetric loudspeaker arrangement \cite{Flux2025SPATRevolution}.  In parallel, object-based delivery standards such as MPEG-H \cite{mpegh2015} reflect a shift from fixed layout assumptions toward a \emph{render-anywhere} philosophy \cite{rumsey2017surround}, with renderers adapting content to a variety of heterogeneous loudspeaker configurations. Recent examples include \emph{Dolby FlexConnect}, a technology which adapts audio content to custom room layouts \cite{flexConnect2024}. Nevertheless, layout-flexible rendering does not obviate the need for layout acquisition: the playback geometry must still be measured or corrected before it can be used by a renderer, which can introduce additional operational overhead for creators \cite{Peters2011CurrentTechnologies}.

\subsection{Interfaces for Spatial Audio Mixing}

Beyond rendering algorithms, prior work has explored more intuitive spatial audio authoring using XR interfaces and has repeatedly identified persistent interaction challenges in 3D audio production. Traditional DAWs and panning tools typically confine 3D spatialization to 2D screen controls, which can impose considerable cognitive burden on sound designers \cite{mathew2017survey, sra_2023, Engeln2025SpatialAudioPaths}. Gelineck and Overholt \cite{gelineck2015haptic} noted that object-based 3D audio production lacks control metaphors, as opposed to stereo production which has one-dimensional controls, or surround production having 2D joysticks and control surfaces. In contrast, AR/VR headsets allow creators to manipulate audio sources directly in the 3D space around them, an approach that prior work has explored as a means of supporting more direct and expressive spatial interaction \cite{Deacon2023SpatialDesign, dearvr2024, Engeln2025SpatialAudioPaths}. A variety of studies have found that creators naturally use hand gestures to describe the spatial motion of sound \cite{Deacon2023Invoke, Deacon2019ShapingSounds}, motivating gesture-based interaction in 6DoF, mapped to the control of audio parameters \cite{Polys2022SpatialAudioDesigner}. For instance, \emph{AudioMiXR} \cite{audiomixr_2025_geleta} lets users wearing a Head-Mounted Display (HMD) grab and position virtual audio objects in their environment, and reports improved usability and creative engagement compared to a conventional 2D panning Graphical User Interface (GUI). Similarly, XR experiences such as \emph{Sonic Sculpture} \cite{martin2020sonic} or \emph{Spatial Orchestra} \cite{kim2024spatial} explore embodied interaction with sound by having participants physically navigate sound-emitting virtual objects in a room or environment. 

However, the majority of XR-based sound design research focuses on head-tracked binaural playback over headphones, rather than reproduction over physical loudspeakers. Even highly immersive AR/VR audio tools often deliver sound through headphones or  the HMD's built-in spatial audio (via head-related transfer functions or HRTFs) for a single user \cite{martin2020sonic, asr_2002, bargum2023spatial, sra_2023}. Prior work \cite{anderson2026grounded} has shown that auditory experiences in XR can benefit from reproduction through physical loudspeakers rather than binaural rendering. While binaural rendering simplifies deployment, it remains an individualized listening experience. Spatial sound installations and performances in the wild typically use loudspeaker systems to envelop \emph{multiple listeners} in a venue. To our knowledge, very few XR authoring interfaces have been deployed in conjunction with \textit{physical} loudspeaker arrays. \emph{ARCube} \cite{arcube2024} is one example of an AR controller paired with a multi-speaker setup, where users manipulate virtual sound objects in a virtual 3D panner cube floating in front of them within a fixed 28-speaker layout. Outside XR, \emph{boomRoom} demonstrates embodied, room-scale manipulation of virtual sound sources reproduced through a custom WFS array and an OptiTrack optical tracking system \cite{the_boomroom_2014}.  These systems demonstrate that embodied spatial interaction and physical loudspeaker reproduction can be combined. Less explored is an authoring workflow in which the loudspeaker geometry itself varies across different venues, is acquired on site, and becomes the coordinate frame for spatial interaction. 

\subsection{Spatial Audio Effects and Perception}

Spatial composition encompasses considerably more than just positioning audio objects in an environment---creators may additionally manipulate their trajectory, diffusion or envelopment (the gradient between localized and diffuse sound), frequency-domain diffusion, spatial granulation, among other techniques \cite{Garcia2017InteractiveCompositional}. The variety is extensive, but equally consequential is understanding how these spatialization strategies are perceived by the listener \cite{Ledoux2018Immersive}. In fact, spatiality is not merely an acoustic property: spatial imagery can arise from memories and imagination \cite{Kendall2010Spatial}. For example, when spatialization is performed across a number of loudspeakers, listeners can perceptually interpolate the gaps in spatial motion and thereby experience continuity \cite{Barrett2012HighOrderAmbisonics}. This perceptual integration can be understood through the lens of Gestalt psychology---the human tendency to organize granular, fragmented events into a coherent whole \cite{Farnell2010DesigningSound}. To illustrate, spatiotemporal illusions are clear examples that perceived spatial behavior need not correspond directly to the discrete acoustic stimulus. Auditory apparent motion \cite{Strybel1992Auditory} and auditory saltation \cite{10.1121/1.423093} are two such illusions: sequential stimuli presented at distinct locations can be perceived as continuous movement or as occupying intermediate positions. Moreover, different spatial dimensions are not perceived with equivalent fidelity. Human spatial hearing is highly anisotropic: left-right localization is more reliable than front-back judgments, while elevation is comparatively inaccurate, and spectral features of the audio source can influence apparent source position, for instance, contributing to illusory elevation \cite{Kendall2010Spatial}. Listeners' head movement can further affect localization by providing additional spatial cues during active listening \cite{Farnell2010DesigningSound}. 

Perception can also be shaped by the audio content being spatialized. Spectral and temporal characteristics of sounds can modulate their affective perception: sharp (high-frequency energy) or rough (high fluctuation in energy) sounds tend to be associated with less pleasant experiences, while lower-frequency, slowly modulating natural ambient sounds are commonly used to evoke relaxation and positive affect \cite{Zhang2022InteractiveSoundDesign}. Musical preferences are also highly idiosyncratic \cite{MCDERMOTT2012227}, with prior work showing that preferences and listening responses can vary with factors such as past experience \cite{Hargreaves1987, Thompson2001MozartEffect}, personality \cite{Litle1986Sensation}, age \cite{LeBlanc1988Tempo}, and gender \cite{McCown1997Bass}. Spatialization can additionally be applied independently to constituent musical layers, raising the question of how different spatial configurations of vocal and instrumental content shape the resulting listening experience. These observations motivate studying spatial rendering as a perceptual configuration involving source trajectory \cite{10.1162/comj_a_00593} (modification of the position parameter), diffuseness or localization \cite{griesinger1997psychoacoustics} (modification of the spread parameter), and audio content, rather than treating geometric source position alone as determinative of listener experience.

\section{System}
The system architecture comprises three principal modules (Fig. \ref{teaser}.B). First, the frontend decoupled into a desktop GUI and an AR controller interface running on a head-worn AR device. The latter renders virtual speaker markers and audio objects, captures user gestures, and supports an in-headset workflow for loudspeaker layout calibration. Second, the \emph{Spatial DJ} backend, running on a host machine, which receives lightweight object-state updates from the frontend, maintains an updated loudspeaker layout which is persisted in the database, and renders the soundscapes (i.e., collections of audio objects) by mapping audio object parameters to per‑speaker gains in real time. The backend also manages the audio stems assigned to each audio object and layout configuration policies (e.g., loudspeakers excluded from the multichannel output). Third, an audio interface routes the multichannel output computed by the backend to a physical loudspeaker array, optionally looping in other intermediaries such as DAWs, controllers, amplifiers, or processors.

\subsection{Frontend Interfaces}

\subsubsection{Desktop GUI}

The \emph{Spatial DJ} desktop GUI (Fig. 2.A) provides an interface for configuring and managing projects,  each of which stores a configurable loudspeaker layout and a collection of audio objects representing monophonic channels. It supports audio source management, including loading audio files and separating tracks into stems using the open-source music source separation model Demucs \cite{rouard2022hybridtransformersmusicsource}, as well as configuration of the loudspeaker layout. For each loudspeaker, users can assign an audio interface output channel, edit the loudspeaker's 3D position of its virtual marker, and calibrate its playback level. Audio sources are represented as virtual audio objects whose spatial rendering parameters can be edited directly from the interface. Users can control each object's 3D position and spread, together with playback states such as mute and solo, and optionally connect external mixer inputs that can be used instead of uploaded audio sources. In addition, a 3D room view provides a live overview of the registered virtual speaker markers and audio objects in the room-aligned frame.

\subsubsection{AR controller}

The AR controller is the immersive counterpart of the frontend that renders the project's virtual speaker markers, registered to physical loudspeakers in the room-aligned coordinate frame (Fig. \ref{teaser}.A), and virtual audio objects. A calibration mode lets the virtual speaker markers be unlocked and supports manual re-alignment by updating their AR world anchors to match the physical layout. 

\begin{figure}[ht!]
    \centering
    \includegraphics[width=1.0\linewidth]{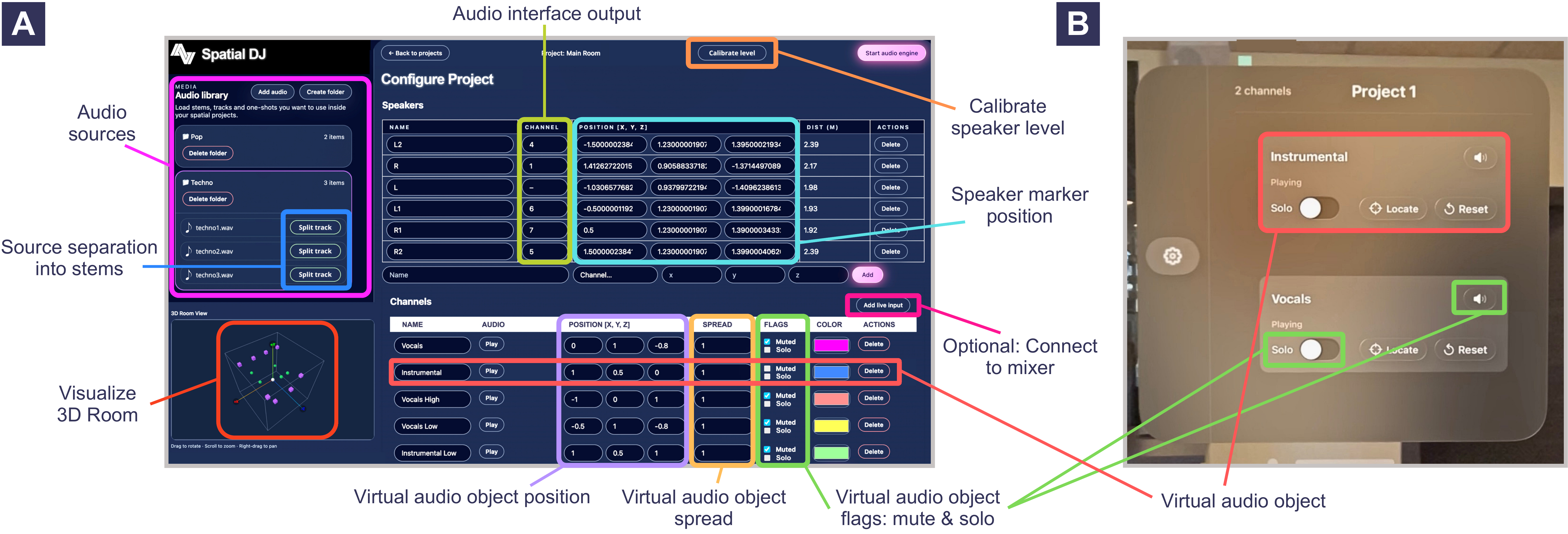}
    \caption{\textbf{\emph{Spatial DJ} frontend UIs}. \textbf{A} The desktop frontend supports project configuration, audio source management and stem separation, loudspeaker layout configuration, speaker level calibration, 3D room visualization, and control of virtual audio object position, spread, flag states, and live mixer inputs. \textbf{B} The AR controller's interface exposes corresponding controls for individual virtual audio objects, enabling users to locate the virtual audio objects in their environment and toggle solo and mute flags.}
    \label{fig:frontend}
\end{figure}

The AR controller captures mid-air gestures (Fig. \ref{teaser}.C--D) such as \emph{drag} (3D translation), \emph{two-hand pinch} (adjustment of the spread parameter, i.e., perceived audio source width), or \emph{single-hand pinch or tap} (mute and solo flags, Fig. \ref{fig:frontend}.B). For communication, the client transmits object-state updates at a cadence of approximately 30 Hz using Open Sound Control (OSC)  over the User Datagram Protocol (UDP). OSC \cite{wright1997osc} is designed for performance and show control contexts---its advantages include interoperability, structured messaging, and low overhead. Each update message carries compact descriptors of object identifiers and parameters, such as position $(x,y,z)$ in the room-aligned frame, spread radius $r$, and state flags. In our implementation, the AR controller runs on an Apple Vision Pro HMD and is implemented in Swift for visionOS 2.0. We use SwiftUI for the application's UI, RealityKit for AR object rendering,  and ARKit \cite{apple_arkit_in_visionos} for world tracking.

\subsubsection{Layout Calibration}
A central design goal is loudspeaker geometry agnosticism: avoiding an assumption of a fixed, symmetric, or predefined loudspeaker arrangement. 
\emph{Spatial DJ} treats calibration as layout acquisition and registration. The system supports two complementary workflows: first, the desktop frontend for entering, editing, or importing loudspeaker positions (per-speaker 3D coordinates) and configuring exclusion policies (Fig. \ref{fig:frontend}.A), and second, the in-headset AR workflow for on-site capture of loudspeaker geometry by initializing an AR session and aligning virtual speaker markers to physical loudspeakers (Fig. \ref{teaser}.D). The resulting 3D loudspeaker layout is persisted and can be reused across sessions by starting from the same origin of the room-aligned coordinate frame.   

\subsection{Backend Audio Processor}\label{sec:backend}

The backend communicates via Object-Relational Mapping (ORM) with the relational database that maintains the up-to-date project data. The backend maps incoming object states from the frontend to per-speaker gains, applies constant power normalization, and mixes audio objects to multichannel output. We differentiate between two audio object types: \emph{point‑source} objects and \emph{ambient} objects. The former are spatialized, i.e., their per‑speaker gains depend on their 3D position and spread parameter, while \emph{ambient} objects are not distance‑spatialized, that is, they are distributed to a selected subset of outputs via a fixed gain vector. The backend is implemented as a Python Flask application exposing a Representational State Transfer (REST) Application Programming Interface (API). The backend runs block-based audio processing---at each audio callback it renders gains from the most recent object state.

\paragraph{Definitions}
Let the physical loudspeakers be indexed by $i=1,...,N$ with room-aligned 3D coordinates $p_{i}=(x_i,y_i,z_i)$. Let the audio objects be indexed by $s\in S$. At each time $t$, let $ S_{\text{pt}}(t) \subseteq S$ and  $S_{\text{amb}}(t) \subseteq S$ denote the objects with \emph{point-source} and \emph{ambient} rendering, respectively, with $S_{\text{pt}}(t) \cap S_{\text{amb}}(t)=\emptyset$. Each object $s$ has a position $p_s(t)=(x_s(t),y_s(t),z_s(t))$, a spread radius $r_s(t)\in[r_{\min},r_{\max}]$, where $r_{\max}$ denotes the maximum supported spread, and a monophonic input signal $x_s[n]$, where $n$ is the sample index. Each loudspeaker $i$ has an output signal $y_i[n]$. Time is $t_n=n/F$, where $F$ is the sampling rate. We also define a set of outputs excluded from point-source normalization,  $E\subseteq \{1,...,N\}$ (e.g., subwoofer). These outputs may still receive fixed gains or separate routing. 

\paragraph{Point-source gain mapping}
For each point-source object $s\in S_{\text{pt}}(t)$ active at time $t$ with position $p_s(t)=(x_s(t),y_s(t),z_s(t))$ and spread radius $r_s(t)<r_{\max}$, we compute the loudspeaker Euclidean-norm distance $d_{i,s}(t)=||p_{i}-p_s(t)||_2$. We first compute raw, nonnegative weights $w_{i,s}(t)$ using a kernel function. In our implementation we use a Gaussian kernel: $w_{i,s}(t)=\exp( - {d_{i,s}^2 (t)/2\sigma^2_s(t)}), i\notin E$.
Here $\sigma_s(t)$ is the kernel width controlling spatial spread. We tie $\sigma_s(t)$ to the user's radius control $r_s(t)$ and clamp to a small minimum value $\sigma_{\text{min}}>0$ to avoid degenerate widths: $\sigma_s(t)=\max(r_s(t), \sigma_{\text{min}})$. We then apply $L_2$
 normalization over the non-excluded outputs so that their squared gain sum remains constant via  $ \varepsilon_s(t) = \sum_{i\notin E} w_{i,s}^2 (t)$. If $\varepsilon_s(t)>0$, gains are $g_{i,s}=w_{i,s}(t)/\sqrt{\varepsilon_s(t)}$, for $i \notin E$, and for excluded outputs $i \in E$ we apply a fixed gain $g_{\text{fixed}}$. 

\paragraph{Ambient objects} When a virtual audio object's spread reaches its maximum value, $r_s(t)=r_{\max}$, it is rendered as an ambient object and is no longer spatialized by distance. Each ambient object is associated with a fixed distribution vector $h_{i,s}$, typically defined over a target speaker subset $\Omega_s\subseteq \{1,...,N\}$. To preserve the squared gain sum independent of the number of selected speakers, we set $h_{i,s}=1/\sqrt{|\Omega_s|}$ for $i\in \Omega_s$ and $h_{i,s}=0$ otherwise, so that $\sum_i h_{i,s}^2=1$.  To avoid an abrupt change at maximum spread, the implementation smoothly blends the normalized point-source gain vector toward the ambient distribution over a small interval below $r_{\max}$ and the width of this transition interval is a hyperparameter. 

\paragraph{Mixing equation}
Before mixing, the backend resolves the mute and solo states of all audio objects into their effective state. If no object is soloed, all non-muted objects remain active. If an object is soloed, only that object remains active. The backend applies envelopes $a_s(t) \in [0,1]$ to smooth the transitions between active ($a_s(t)=1$) and inactive ($a_s(t)=0$) states to avoid abrupt level changes. The multichannel output is then mixed as (Eq. \ref{eq:mixing-equation}):
\begin{equation}
    y_i[n] =  \sum_{s\in S_{\text{pt}}}a_s(t_n)g_{i,s}(t_n)x_s[n] + \sum_{s\in S_{\text{amb}} } a_s(t_n) h_{i,s} x_s[n]
    \label{eq:mixing-equation}
\end{equation}

\subsection{Audio Interface and Output Chain}
\emph{Spatial DJ} uses a low-latency Audio-over-IP (AoIP) multichannel input/output path between the backend and the loudspeaker array, with one discrete output channel per loudspeaker (i.e., a one-to-one mapping from output channels to loudspeakers)  with sufficient channel count for the target loudspeaker array, and support stable sample rate and buffer size for real-time rendering.

The system supports both active and passive loudspeaker deployments. Passive systems require external amplification, while active loudspeakers integrate amplification per cabinet. External processors and playback tools (e.g., a DJ controller or DAW) may be incorporated upstream for asset playback and global equalization or level tuning, whereas the \emph{Spatial DJ} backend continues to control the spatial rendering parameters of each audio object. 

\section{Technical Evaluation}
\begin{figure*}[ht!]
    \centering
    \includegraphics[width=1.0\linewidth]{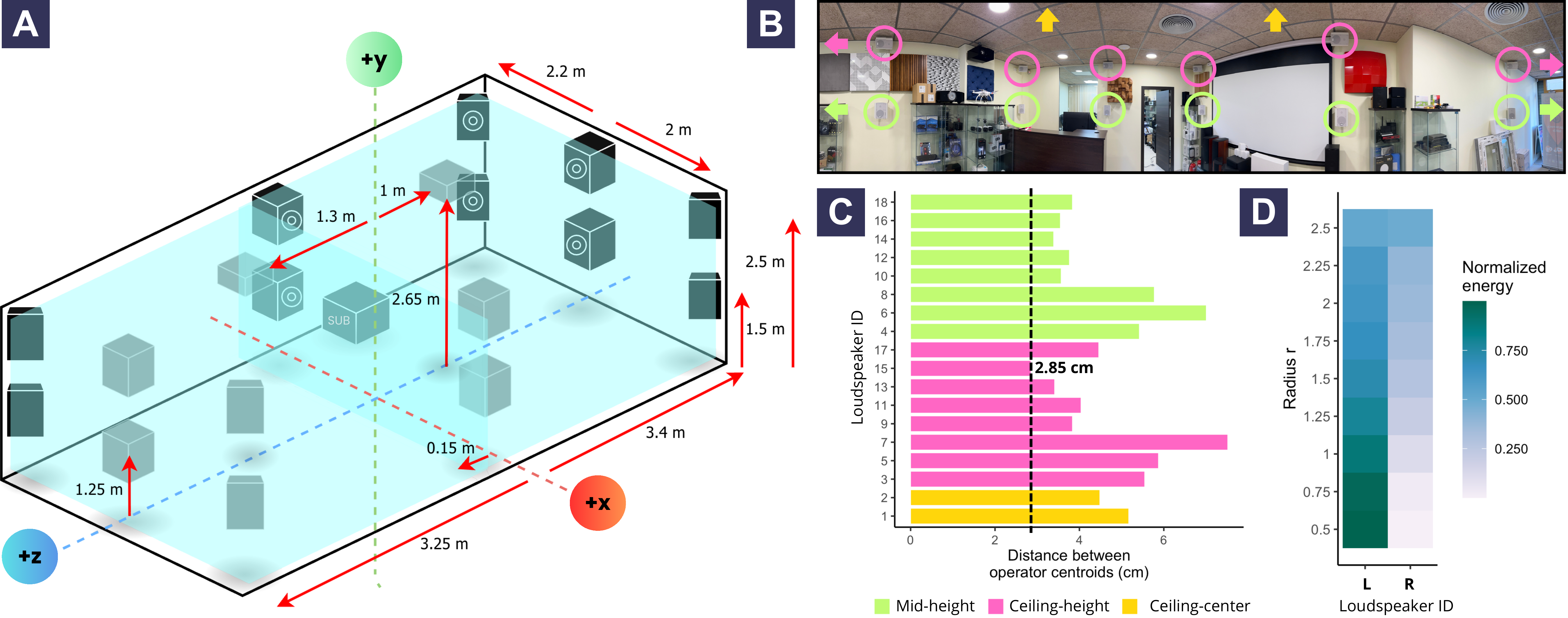}
    \caption{\textbf{Controlled evaluation setup and key measurements}. \textbf{A} Schematic of the 18-loudspeaker room with arrows indicating the measured distances from the room axes to the loudspeaker position coordinates. \textbf{B} Panoramic view with a subset of loudspeaker locations,  arrows indicate additional speakers outside the field of view. \textbf{C} Between-operator calibration discrepancy per loudspeaker. The dashed line marks the smallest between-operator discrepancy. \textbf{D} Normalized per-speaker energy as the radius parameter increases, illustrating the transition from point-source to ambient rendering in a stereo setup.}
    \label{fig:radius-transition}
\end{figure*}

To address RQ$_1$, we evaluate the system in a controlled test environment: a room measuring approximately 4.46\,m $\times$ 2.58\,m $\times$ 6.96\,m configured with $N{=}$18 loudspeakers at fixed positions (Fig. \ref{fig:radius-transition}.A and \ref{fig:radius-transition}.B). The loudspeaker layout comprises a perimeter array with two height layers (8 speakers at mid-height and 8 speakers at ceiling height), plus a ceiling cluster of two speakers near the room center. We use a passive-speaker deployment, with compact Class-D stereo amplifiers with balanced inputs to drive the 18-loudspeaker group of small passive cabinets (of size 24.8\,cm $\times$ 18.3\,cm $\times$ 16.4\,cm, with 5-inch woofer drivers, and nominal 8~$\Omega$\,impedance). The playback pipeline routes through a multichannel AoIP interface into the \emph{Spatial DJ} backend. Before the evaluation, we performed level calibration by reproducing the same white noise stimulus through each loudspeaker individually and adjusting its output gain until all loudspeakers produced the same measured dB SPL at a fixed origin marker at the room center, using a UNI-T UT353 Sound Level Reader Decibel Meter. 

We define a right-handed room coordinate system with its origin at a floor marker placed at the geometric center of the room: $+x$ is aligned with the room's shorter horizontal dimension (the 4.46\,m span), $+y$ is vertical, and $+z$ is aligned with the orthogonal longer horizontal dimension (the 6.96\,m span). Ground-truth speaker positions are measured in this frame using a laser distance meter and measuring tape, yielding $(x,y,z)$ coordinates per speaker. To reduce measurement noise, each speaker position is measured three times and averaged. 

\subsection{Layout Calibration Accuracy}
We evaluate how accurately the AR calibration procedure recovers known loudspeaker positions and how repeatable the resulting estimates are across different operators and repeated trials. Each trial begins from a clean calibration state. The operator calibrates all $N{=}18$ speakers using the AR controller by switching to the calibration mode and unlocking the virtual speaker markers. Then, the operator selects a virtual speaker marker, aligns it with the center of the loudspeaker woofer, and confirms the new position. This process continues until all loudspeakers are registered. For each loudspeaker $i$, trial $k$, and operator $o$, we record the estimated position $\hat{\mathbf{p}}_{i,k,o}\in\mathbb{R}^3$. We perform 10 full calibrations per operator (Op$_A$: $k{=}1..10$, Op$_B$: $k{=}1..10$), yielding a total of 360 calibrated speaker marker positions.

Let $\mathbf{p}_i$ denote the ground-truth position of loudspeaker $i$ and $\hat{\mathbf{p}}_{i,k,o}$ the calibrated estimate. We quantify accuracy using the Euclidean error $e_{i,k,o}=\lVert \hat{\mathbf{p}}_{i,k,o}-\mathbf{p}_i\rVert_2$ (cm) and additionally report axis-wise absolute deviations $|\Delta x|$, $|\Delta y|$, and $|\Delta z|$ to characterize the directional distribution of calibration error. Across all speakers and trials, the Euclidean error has median $5.51$\,cm (IQR 3.62--8.78\,cm), with 95\% bootstrap CI of [4.49, 7.26]\,cm using resampling over trials ($B{=}5{,}000$ resamples). Axis-wise absolute deviations are $|\Delta x|{=}4.10$\,cm (IQR 1.92--6.68\,cm, 95\% CI [3.13, 5.07]\,cm), $|\Delta y|{=}0.59$\,cm (IQR 0.24--0.98\,cm, 95\% CI [0.55, 0.64]\,cm), and $|\Delta z|=2.51$\,cm (IQR 1.07--4.41\,cm, 95\% CI [1.94, 3.35]\,cm). 

Across 10 repeated full-room AR calibrations, the median deviation from each operator's per-speaker centroid was 5.01 cm for Op$_A$ and 4.20 cm for Op$_B$. Tail deviations remained modest, with p95${=}14.9$ cm (Op$_A$) and 19.2 cm  (Op$_B$). These deviations should also be interpreted relative to the physical dimensions of the loudspeakers used in the evaluation. Small differences in marker placement across trials could therefore correspond to slightly different positions on the same loudspeaker cabinet or around the woofer. Axis-wise variation showed an anisotropic pattern: the median absolute deviation was smallest in height (0.578 cm for Op$_A$ and 0.410 cm for Op$_B$), suggesting that vertical placement was more repeatable than placement along the horizontal axes. In contrast, the dominant variation occurred in the horizontal plane, particularly along the room $x$-axis (3.54 cm Op$_A$ and 2.21 cm Op$_B$) and $z$-axis (2.89 cm Op$_A$ and 2.30 Op$_B$), consistent with alignment uncertainty when registering perimeter speakers. Comparing operator centroids (Fig. \ref{fig:radius-transition}.C), the median between-operator discrepancy was 4.25 cm with p95${=}7.08$ cm, a maximum of 7.52 cm, and a minimum of 2.85 cm, suggesting limited operator-dependent bias and that independently calibrated layouts converge to essentially the same speaker geometry. 

\subsection{Control Latency}
\begin{figure}[ht!]
    \centering
    \includegraphics[width=0.75\linewidth]{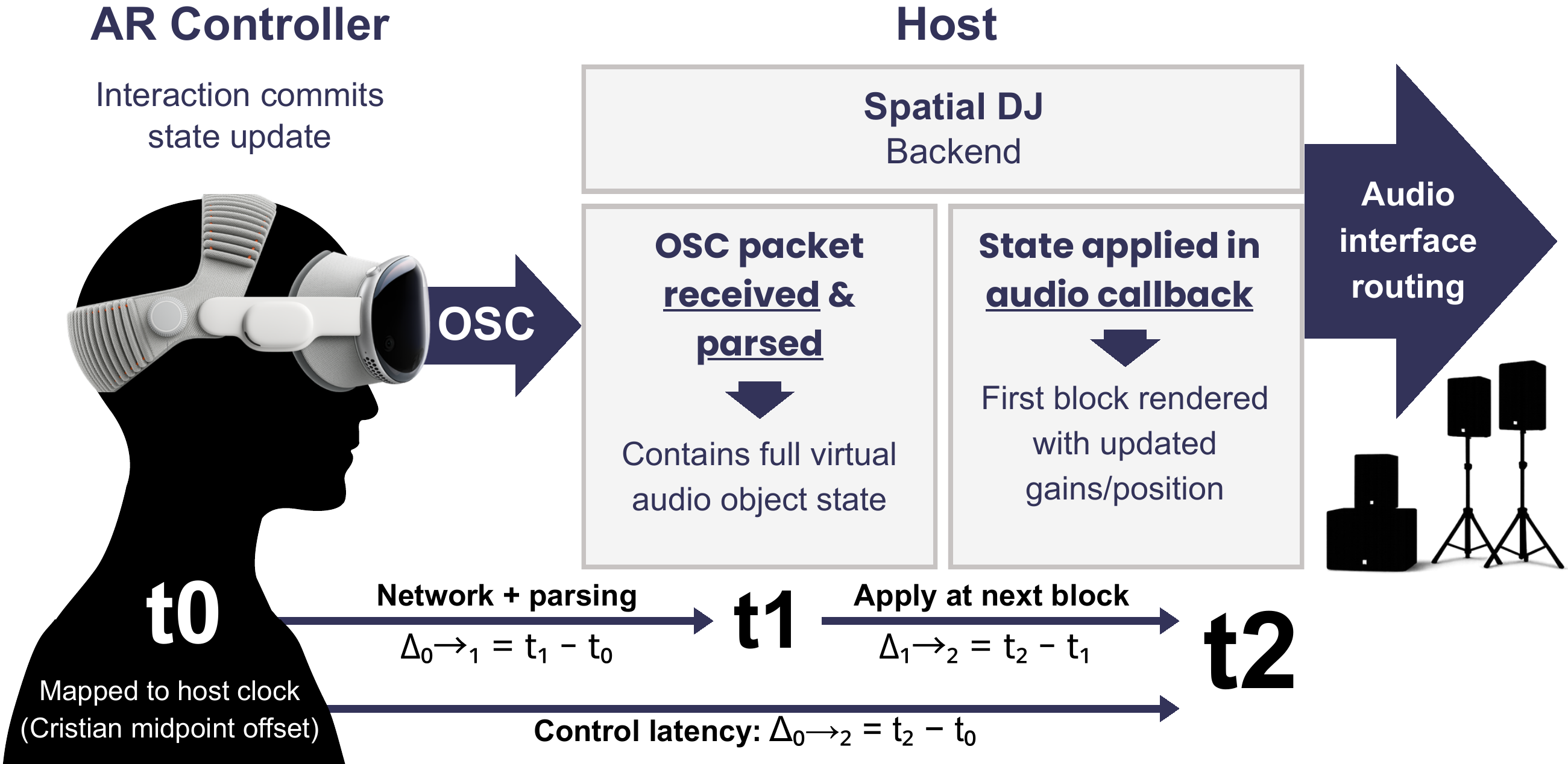}
    \caption{\textbf{Control latency measurement pipeline}. The AR controller records $t_0$ when an interaction commits an object-state update and transmits it via OSC. Then, the host records $t_1$ when the packet is received and parsed, and $t_2$ when the update is first applied inside the audio callback.}
    \label{fig:times}
\end{figure}

We define control latency as the delay from committing an object-state update on the AR controller to that update first being applied by the audio rendering callback. This measurement excludes gesture recognition latency. We characterize how this delay varies with scene complexity. Latency is measured using three timestamps corresponding to distinct stages in the interaction pipeline (Fig. \ref{fig:times}): $t_0$ is recorded on the AR controller at the moment an interaction commits a state update, immediately prior to transmitting the OSC message; $t_1$ is recorded on the host when the OSC packet is received and parsed, and $t_2$ is recorded in the audio engine when the updated state is first applied inside the audio callback, operationalized as the first output block rendered with the updated gains.

Because $t_0$ originates on the AR controller while $t_1$ and $t_2$ originate on the backend host, we map controller timestamps into the host clock domain using an offset-estimation procedure based on Cristian's algorithm \cite{cristian1989probabilistic}. Specifically, prior to each run the controller queries a lightweight host endpoint returning the host time in nanoseconds and, using a round-trip midpoint estimate, the controller computes a clock offset and records $t_0$ in \emph{host-time} as $t_0^{(\mathrm{host})}=t_0^{(\mathrm{client})}+\hat{\Delta}$, which allows meaningful computation of $\Delta_{0\to1}=t_1-t_0$ and $\Delta_{0\to2}=t_2-t_0$.

From these measurements we compute the total control latency $\Delta_{0\to2}=t_2-t_0$, and decompose it into $\Delta_{0\to1}$, capturing network transport and message parsing, and $\Delta_{1\to2}$, capturing engine scheduling and audio-block quantization effects. For each condition, the operator continuously moved the active virtual sound objects throughout the listening area while the client streamed OSC updates at a target rate of approximately $30$\,Hz per object for $10$\,s. In practice, this yielded roughly $250$ updates per object per run. Each OSC update includes a unique message identifier and the backend logs a single latency measurement for an update when that identifier is first observed inside the audio callback.

\begin{table}[ht!]
  \caption{Control latency (ms) for block size $B=256$ at 44.1\,kHz. Values report median [p25, p75], mean (SD), and p95.}
  \label{tab:latency}
  \centering
  \begin{tabular}{lrrrr}
    \toprule
    \textbf{Metric} & \textbf{$n$} & \textbf{Median [p25, p75]} & \textbf{Mean (SD)} & \textbf{p95} \\
    \midrule

    $\Delta_{0\to1}$ (network+parse) & 6164 & 2.18 [1.30, 3.18] & 4.55 (9.38) & 22.54 \\
    $\Delta_{1\to2}$ (engine+block)  & 6164 & 5.15 [2.63, 7.87] & 5.33 (4.52) & 9.82 \\
    \textbf{$\Delta_{0\to2}$ (control)} &
    \textbf{6164} &
    \textbf{8.07 [5.07, 11.00]} &
    \textbf{9.88 (10.08)} &
    \textbf{27.40} \\

    \addlinespace[0.6em]
    \multicolumn{5}{l}{\textbf{$\Delta_{0\to2}$ by number of active speakers $N$}} \\
    $N=2$  & 1596 & 8.20 [5.34, 11.04] & 9.96 (8.52)   & 27.89 \\
    $N=4$  & 1629 & 8.11 [4.96, 10.73] & 9.01 (6.89)   & 22.36 \\
    $N=8$  & 1403 & 7.72 [4.75, 11.01] & 10.58 (13.11) & 31.43 \\
    $N=18$ & 1536 & 8.21 [5.28, 11.28] & 10.08 (11.14) & 24.77 \\

    \addlinespace[0.6em]
    \multicolumn{5}{l}{\textbf{$\Delta_{0\to2}$ by number of active objects $S$}} \\
    $S=1$ & 881  & 8.55 [5.43, 11.76] & 10.64 (12.19) & 27.07 \\
    $S=2$ & 1627 & 7.80 [4.79, 10.76] & 9.66 (8.69)   & 28.55 \\
    $S=4$ & 3656 & 8.08 [5.09, 10.97] & 9.79 (10.09)  & 25.75 \\

    \bottomrule
  \end{tabular}
\end{table}

Table~\ref{tab:latency} summarizes the empirical latency distributions over $n{=}6{,}164$ applied updates. Control latency is low: median $\Delta_{0\to2}=8.07$\,ms with an interquartile range (IQR) of $[5.07,\,11.00]$\,ms and a 95th percentile of $27.40$\,ms. Decomposition shows that $\Delta_{1\to2}$ contributes the largest share (median $5.15$\,ms), consistent with audio-block quantization (256 frames at 44.1\,kHz corresponds to 5.80\,ms), while network+parsing $\Delta_{0\to1}$ had a lower median ($2.18$\,ms) but exhibits heavier tails. Within these benchmark runs, median control latency varied little across the scene complexity range. Increasing the number of active speakers from $N{\in}\{2,4,8,18\}$ changes the median control latency by at most $0.49$\,ms (max--min across $N$), and increasing the number of concurrently rendered objects from $S{\in}\{1,2,4\}$ changes the median by at most $0.74$\,ms (max--min across $S$). Within the evaluated configurations, these results suggest that the overall delay is dominated by network transport or packet handling and audio callback scheduling rather than the number of active sources or speakers for the evaluated configurations.

\subsection{Spread Parameter Validation}
We validate that the spread radius parameter $r$ produces a controllable transition from point-source rendering (energy concentrated near the closest speakers) to ambient rendering (i.e., diffuse, approximately uniform distribution across loudspeakers). In the implementation, the normalized point-source gain distribution is smoothly blended toward the ambient distribution over the final 0.2 units of the supported radius range, with $r_{\text{max}}=2.5$ corresponding to ambient rendering. We use a stereo setup with two anchors ($N=2$): a left and a right loudspeaker. We then place a single virtual audio object at a fixed position $\mathbf{p_\text{obj}}$ coincident with the left loudspeaker position. We render a stationary white noise stimulus at $-6$\,dBFS and sweep $r$ over the backend's supported range $r \in [0.5, 2.5]$ in steps of $\nicefrac{1}{4}$. For each radius value, we render 30\,s of audio and record the multichannel output digitally (one channel per loudspeaker). For each analysis window $t$, we compute per-speaker RMS amplitude  $E_{i,t}(r)$ of each loudspeaker channel $i$. For each radius $r$, the set $\mathbf{E}(r)$ contains all per-channel RMS observations recorded across the analysis windows. We then compute the coefficient of variation $\mathrm{CV}(r) = \mathrm{std}(\mathbf{E}(r))/\mathrm{mean}(\mathbf{E}(r))$ and the peak to median ratio $\max{(\mathbf{E}(r))}/\text{median}(\mathbf{E}(r))$, where lower values indicate less variation among the recorded channel RMS amplitudes. 

As $r$ increases from 0.5 to 2.5, the output becomes progressively less concentrated in the nearest loudspeaker (Fig. \ref{fig:radius-transition}.D): the peak to median ratio drops from 2.016 at $r{=}0.5$ to 1.033 at $r{=}2.5$, and uniformity improves (CV from 1.008 to 0.029). At $r=2.5$, the backend renders the audio object  as ambient. These results validate that the spread parameter provides a controllable transition from concentrated point-source rendering to an ambient distribution across loudspeakers at maximum radius set as a hyperparameter.

\section{Controlled Listener Study}

In relation to RQ$_2$, we conducted a controlled listener study using \emph{Spatial DJ} to compare different spatial rendering configurations (i.e., modifying position and spread parameters) applied to vocal and instrumental audio objects. The study was conducted in the same 18-loudspeaker environment employed in the technical evaluation (Fig. \ref{fig:radius-transition}.B). We recruited 18 participants (10 male, 8 female) with self-reported normal hearing using snowball recruitment during May 2026 and August 2026 in Spain. 

All participants provided informed consent and took part voluntarily. The aim of the study was to characterize how listeners experienced spatial behaviors supported by \emph{Spatial DJ}. For experimental consistency, rather than manually authoring the motion during each trial, we used predefined rendering configurations so that every participant experienced the same programmed behavior under each condition while still reflecting the kinds of trajectories that could be created through free-hand authoring. 

\paragraph{Stimuli and Conditions} We selected four 30-second musical excerpts, denoted as M$_1$--M$_4$, representing Pop, House, Hip Hop, and Techno genres, to provide variation in musical content. Each excerpt was separated into vocal and instrumental stems in the desktop GUI. We independently manipulated the position and spread parameters of the vocal and instrumental stems, rendered as individual audio objects in \emph{Spatial DJ}. Each one was rendered either as \emph{ambient}, distributed equally across all loudspeakers, or as a \emph{point-source} object following an automated elliptical orbit. For each excerpt, we selected an angular velocity through manual perceptual tuning, and kept this velocity constant across all orbiting conditions for that excerpt. Expressing the selected velocities relative to tempo gives an angular displacement of approximately $22$--$30^\circ$ per beat, equivalent to one revolution every $12$--$16$ beats. 

\begin{figure}[ht!]
    \centering
    \includegraphics[width=\linewidth]{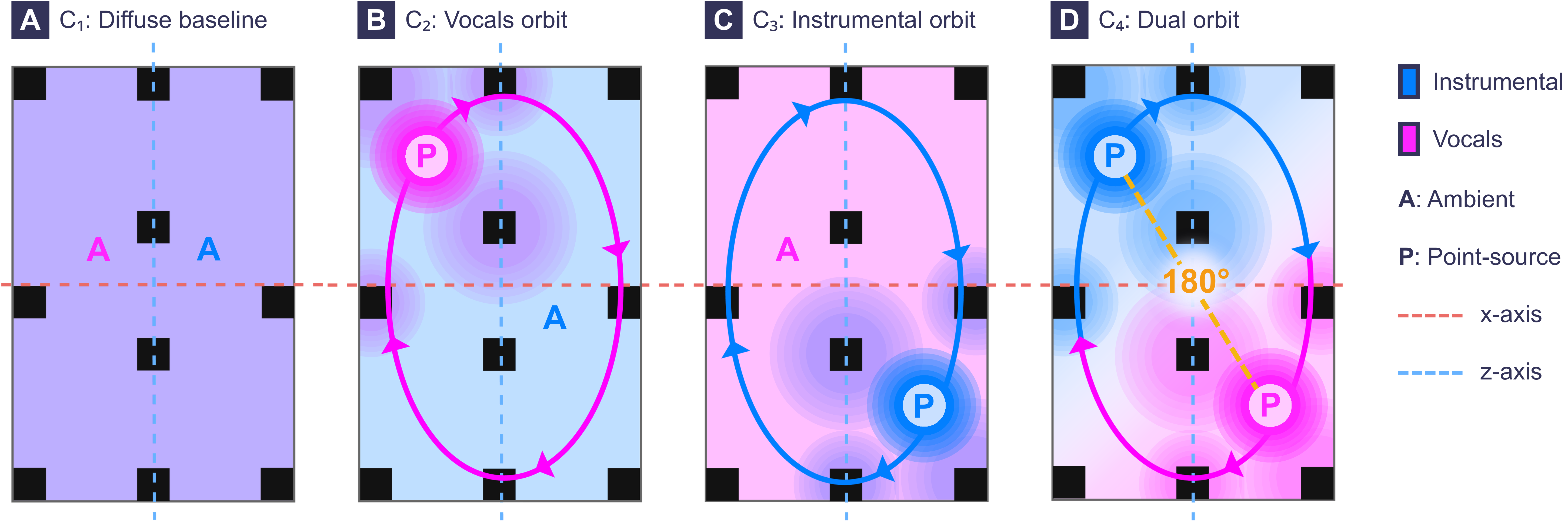}
    \caption{\textbf{Spatial audio conditions used in the listener study, illustrated in a representative top view of the 18-loudspeaker room where the study was conducted}. \textbf{A}  C$_1$, diffuse baseline: the vocal and instrumental layers are rendered as ambient audio objects.
    \textbf{B}  C$_2$, vocal orbit: the vocal layer follows a clockwise elliptical orbit as a point-source object while the instrumental layer remains ambient.
     \textbf{C}  C$_3$, instrumental orbit: the instrumental layer follows a clockwise elliptical orbit as a point-source object while the vocal layer remains ambient.
      \textbf{D}  C$_4$, dual orbit: the vocal and instrumental layers follow the same clockwise elliptical orbit with the same angular velocity and a $180^\circ$ phase offset.
      Blue represents the instrumental layer and pink represents the vocal layer, while A and P denote ambient and point-source rendering, respectively.}
    \label{fig:conditions}
\end{figure}

Combining the two object types resulted in four conditions (Fig. \ref{fig:conditions}): a diffuse baseline (C$_1$), vocal orbit (C$_2$), instrumental orbit (C$_3$), and dual orbit (C$_4$). In the dual orbit condition, both audio objects followed elliptical orbits at the same angular velocity, with a $180^\circ$ phase offset.

Playback levels were adjusted separately for each excerpt and condition to minimize loudness differences between trials. Sound level was measured at the marked listening position using UNI-T UT353 Sound Level Reader Decibel Meter. Playback levels fluctuated between 60 and 70 dBA over the course of each excerpt, and no excerpt exceeded 70 dBA.  

\paragraph{Experimental Design} We used a repeated-measures design for vocal $\times$ instrumental rendering configurations, such that every participant experienced all four conditions. To limit the duration of the study while incorporating variation in musical content, each participant was assigned two of the four excerpts and heard each excerpt once under every condition, therefore resulting in eight trials per participant. Trials were organized into two four-trial blocks, with one musical excerpt used throughout each block. We constructed 18 participant schedules to counterbalance condition (C$_1$--C$_4$) and the presentation of musical excerpts (M$_1$--M$_4$). The schedules were sampled without replacement from the 24 possible permutations of the four spatial conditions and selected to distribute each condition as evenly as possible across within-block presentation positions. Musical excerpts were similarly distributed across participants and between the first and second blocks. 

\paragraph{Procedure} On arrival, participants reviewed the study information, provided informed consent, and completed eligibility screening for self-reported hearing. Participants then received instructions explaining that they would hear musical excerpts presented with different spatial effects, without being informed of the condition labels or intended spatial trajectories. Participants were then seated at a marked listening position at the center of the room. Each participant completed eight trials organized into two four-trial blocks. In each trial, they listened to a 30-second excerpt and immediately afterward completed the questionnaire about their listening experience. A short break was offered between blocks, and participants could pause or discontinue the study at any time. On average, a complete session lasted approximately 15--20 minutes per participant, including instructions, questionnaires, and breaks.

\paragraph{Measures} After each trial, participants rated six statements assessing multiple exploratory dimensions of their audio experience: perceived immersion, motion clarity, vocal focus, energy, novelty, and distraction or discomfort, on a seven-point Likert scale. Participants also answered a categorical question indicating which type of spatial motion they perceived, if any. Open-ended responses were used illustratively to contextualize the quantitative ratings. The complete questionnaire is provided in Appendix \ref{appendix}. 

\subsection{Results}
We analyzed each perceptual outcome using a linear mixed-effects model with spatial condition (C$_1$--C$_4$), music block (1 or 2), within-block presentation position (1--4), and musical excerpt (M$_1$--M$_4$) as fixed effects, and participant as a random intercept. Estimated marginal means (EMMs) were used to characterize condition effects after adjustment for excerpt choice and presentation order. Our primary contrast compared the perceived immersion between the diffuse baseline (C$_1$) and vocal orbit condition (C$_2$). Secondary pairwise comparisons were Tukey-adjusted within each outcome. Below, values following $\pm$ indicate the half-width of the 95\% confidence interval (CI). Neither music block nor position within the four-trial block significantly affected any of the six outcomes. 

Participants' categorical reports of apparent motion (Table \ref{tab:motion-perception}) reveal that it was not a bijective mapping with the programmed source trajectory: C$_2$--C$_4$ were predominantly recognized as moving and elliptical, while in C$_4$ listeners detected the increased structural complexity and was predominantly associated with two moving layers; however, the exact motion description varied across listeners. C$_1$ was the most spatially ambiguous as changes in the spectral distribution of the musical signal may have contributed to momentary sensations of apparent motion.   

\begin{table}[th!]
\centering
\caption{\textbf{Perceived motion types by spatial condition}. Values indicate the number of trials and percentage of trials within each condition ($N=144$ trials, 18 participants $\times$ 8 trials, 36 observations per condition). Participants could select more than one motion type per trial, so percentages within a condition do not sum to 100\%.}
\label{tab:motion-perception}
\begin{tabular}{lcccc}
\toprule
\textbf{Perceived motion type} &
\textbf{C$_1$} &
\textbf{C$_2$} &
\textbf{C$_3$} &
\textbf{C$_4$} \\
\midrule
Diffuse or scattered
& \textbf{16 (44.4\%)}
& 7 (19.4\%)
& 10 (27.8\%)
& 10 (27.8\%) \\

Elliptical movement
& 2 (5.6\%)
& \textbf{26 (72.2\%)}
& \textbf{28 (77.8\%)}
& \textbf{25 (69.4\%)} \\

Horizontal movement
& 8 (22.2\%)
& 9 (25.0\%)
& \textbf{18 (50.0\%)}
& 8 (22.2\%) \\

Vertical movement
& 7 (19.4\%)
& 1 (2.8\%)
& 0 (0.0\%)
& 1 (2.8\%) \\

Two moving layers or objects
& 5 (13.9\%)
& 9 (25.0\%)
& 9 (25.0\%)
& \textbf{23 (63.9\%)} \\

Not sure
& \textbf{5 (13.9\%)}
& 0 (0.0\%)
& 0 (0.0\%)
& 0 (0.0\%) \\
\bottomrule
\end{tabular}
\end{table}

\begin{figure}[ht!]
    \centering
    \includegraphics[width=\linewidth]{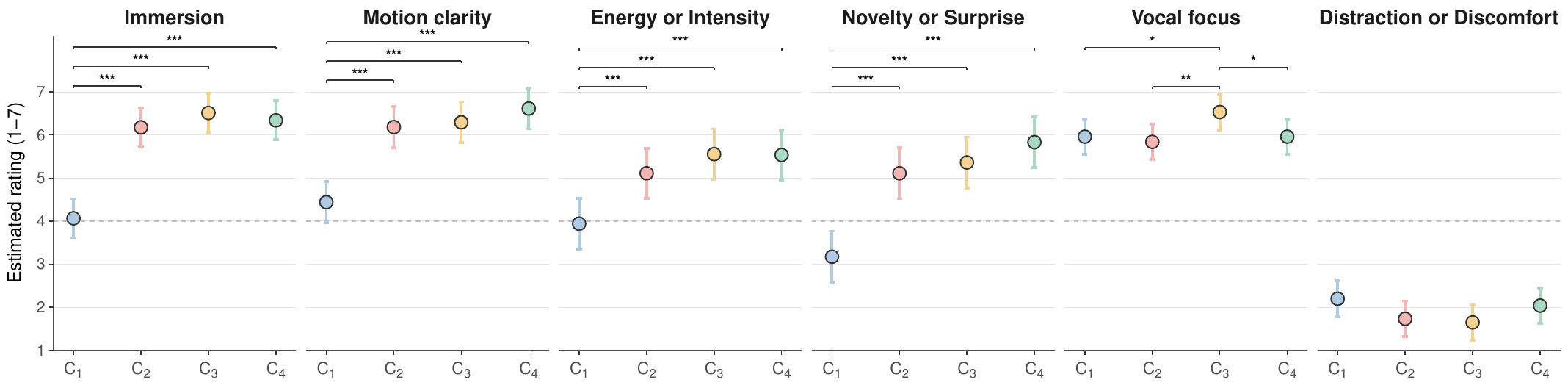}
    \caption{\textbf{Model-adjusted perceptual ratings across study conditions}. Points show estimated marginal means and vertical error bars indicate 95\% confidence intervals, adjusted for musical excerpt, music block, and within-block presentation position. C$_1$: Diffuse baseline; C$_2$: Vocal orbit; C$_3$: Instrumental orbit; C$_4$: Dual orbit. (*$p <.05$, **$p<.01$, ***$p<.001$.)}
    \label{fig:results-bars}
\end{figure}

\paragraph{Orbiting point-source conditions received higher immersion ratings than the diffuse baseline} Study condition had a statistically significant effect on immersion,  $F(3, 116.31)=35.26, p<.001$. The diffuse baseline received the lowest estimated immersion rating (C$_1$: $\mu_{\text{EMM}}=4.06\pm 0.45$, 95\% CI), compared with each of the moving conditions: C$_2$ ($\mu_{\text{EMM}}=6.18\pm 0.45$), C$_3$ ($\mu_{\text{EMM}}=6.51\pm 0.45$), and C$_4$ ($\mu_{\text{EMM}}=6.34\pm 0.45$). The primary contrast C$_2-$C$_1$ showed that moving the point-source vocals increased immersion by $2.11\pm 0.54$ points on the seven-point scale, with 95\% CI, $t(116.31)=7.74, p<.001$ (Fig. \ref{fig:results-bars}). All Tukey-adjusted pairwise comparisons against C$_1$ reached $p<.001$, with estimated differences of  $2.11\pm 0.71$ points for C$_2$, $2.45\pm 0.71$ for C$_3$, and $2.28\pm 0.71$ for C$_4$. P03 explicitly connected spatial effects to immersion and attentional tracking: \enquote{\emph{Immersion changes it [the perception], when different sounds come from different sides, you perceive them differently and instinctively follow them}}, P04 gave a direct moving versus static comparison: \enquote{\emph{I did not like the version without the effect at all---it sounded muffled---whereas I liked the version with effects}}, and P05 and P06 also referred to dancing: \enquote{\emph{I liked it with the effects. Different effects made me want to dance in different ways}} and \enquote{\emph{the memorable part was the elliptical movement [...] it would be great for dancing or a club}}. 

\paragraph{The dual orbit condition did not improve motion clarity over single orbit conditions} In the study, motion clarity was lowest in C$_1$ ($\mu_{\text{EMM}}=4.44\pm 0.48$, 95\% CI), higher in the vocal orbit (C$_2$: $\mu_{\text{EMM}}=6.18\pm 0.48$, 95\% CI) and background orbit conditions (C$_3$: $\mu_{\text{EMM}}=6.30\pm 0.48$, 95\% CI), and highest in the dual orbit condition (C$_4$: $\mu_{\text{EMM}}=6.61\pm 0.48$, 95\% CI). Tukey-adjusted comparisons showed that C$_4$ exceeded C$_1$ by 2.18 points, $t(115.64)=7.79, p<.001$ (Fig. \ref{fig:results-bars}). However, despite having the highest estimated mean, C$_4$ did not differ significantly from C$_2, p=.414$, or C$_3, p=.661$. C$_2$ and C$_3$ also did not differ, $p=.978$.    

\paragraph{Vocal focus ratings were highest in the instrumental orbit condition} Vocal focus ratings were high across all conditions (C$_1$: $\mu_{\text{EMM}}=5.96\pm 0.41$, 95\% CI; C$_2$: $\mu_{\text{EMM}}=5.84\pm 0.41$, 95\% CI; C$_3$: $\mu_{\text{EMM}}=6.53\pm 0.41$, 95\% CI; and C$_4$: $\mu_{\text{EMM}}=5.96\pm 0.41$, 95\% CI). However, our Tukey-adjusted comparisons showed that C$_3$ produced higher vocal focus ratings than C$_1$, with $\Delta=0.57 \pm 0.56, p=.044$, more than C$_4$, with $\Delta=0.57 \pm 0.56, p=.043$, and even the vocal orbit condition C$_2$, with $\Delta=0.69 \pm 0.56, p=.009$. Qualitatively, four participants specifically mentioned their favorite condition was C$_3$, P18 said \enquote{\emph{it was the most memorable. I wanted to keep listening}}. 

\paragraph{Orbiting point-source conditions received higher energy and novelty ratings without a detectable discomfort penalty} Orbiting conditions made the excerpts feel significantly more energetic or intense (Tukey-adjusted comparisons showed that C$_2$ exceeded C$_1$ by $1.17\pm 0.78$ points, 95\% CI, $p<.001$, while C$_3$ and C$_4$ surpassed C$_1$ with $1.62 \pm 0.78$ and $1.60\pm 0.78$, respectively, all $p<.001$), but coordinated movement of two audio objects did not produce a significantly detectable advantage over simpler single-object motion (Fig. \ref{fig:results-bars}). P07 connected motion to both spaciousness and energy: \enquote{\emph{it is flatness versus spaciousness---the moving effects add more energy}}. The diffuse baseline C$_1$ was rated least novel ($\mu_{\text{EMM}}=3.17\pm 0.59$, 95\% CI) while C$_4$ was descriptively the most novel condition ($\mu_{\text{EMM}}=5.83\pm 0.59$, 95\% CI). P11 wrote: \enquote{\emph{it was interesting when the voice and music moved at the same time}}. However, its difference from C$_2$, whose $\mu_{\text{EMM}}=5.11\pm 0.59$, 95\% CI, was not significant after Tukey adjustment, $p=.104$. Similarly, C$_4$ also did not differ significantly from C$_3$ ($\mu_{\text{EMM}}=5.36\pm 0.59$, 95\% CI), with $p=.439$. Finally, the overall effect of a condition on distraction or discomfort did not reach significance, $F(3, 115.51)=2.32, p=.079$. Ratings remained low across all conditions and no Tukey-adjusted comparison was significant, all $p\geq.105$. Qualitatively, dual source motion was highly unusual and memorable, as P07 said \enquote{\emph{separating the vocal and melody---when they move in parallel---is cool. I have never heard anything like this before}}, but also polarizing, described as excessive by some participants, if played for a prolonged time: P08 said \enquote{\emph{with two moving objects, you could get dizzy... If it continued for a long time, I would go crazy}}, P10: \enquote{\emph{when both sounds were moving, it was too much}}, and P12: \enquote{\emph{it would be good in a club, like a stroboscopic effect, for a few seconds, but not continuously}}. 

\section{Representative Deployments}\label{sec:experiments}

\begin{figure*}[ht!]
    \centering
    \includegraphics[width=\linewidth]{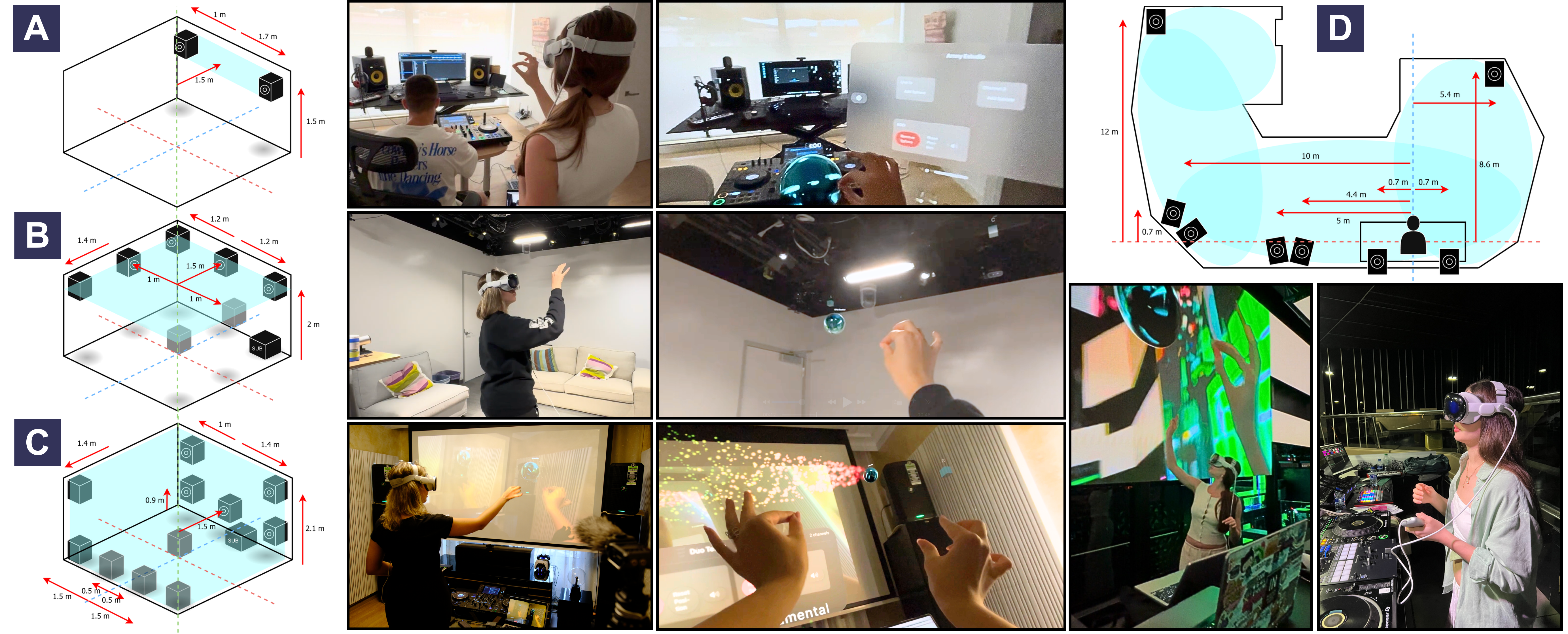}
    \caption{\textbf{Representative deployments across heterogeneous loudspeaker configurations}. \emph{Left for A--C, top for D}: schematic layouts (not to scale) showing loudspeaker placement and room dimensions for four tested setups: \textbf{A} 2.0 stereo baseline with two active loudspeakers. \textbf{B} 7.1 laboratory surround system routed through an A/V receiver (planar layout). \textbf{C} 7.1.4 home-cinema system with passive loudspeakers (periphonic layout with height channels). And \textbf{D} nightclub venue with eight non-uniformly spaced loudspeakers and an asymmetric performer/DJ position. \emph{Right for A--C, bottom for D}: photographs from each deployment showing on-site spatial audio control with the AR interface in the corresponding physical space.}
    \label{fig:setups}
\end{figure*}

To complement our controlled evaluations, we report a set of representative deployments of \emph{Spatial DJ} that stress the system across heterogeneous loudspeaker configurations and deployment contexts: a 2.0 stereo baseline with two active loudspeakers (Fig. \ref{fig:setups}.A); a 7.1 surround sound lab setup routed through a consumer A/V receiver Sony ES STR-AZ1000ES (Fig. \ref{fig:setups}.B); a 7.1.4 periphonic sound system in a home-cinema equipped with passive loudspeakers and external amplification, routed through a multichannel preamplifier processor Yamaha Natural Sound CX-A5100 feeding a multichannel power amplifier Yamaha Natural Sound MX-A5000 (Fig. \ref{fig:setups}.C), 
and a nightclub venue with eight non-uniformly spaced loudspeakers positioned asymmetrically relative to the DJ booth, with professional stereo power amplifiers Crown XLS 5000 and connected to DJM-S11 mixer (Fig. \ref{fig:setups}.D).

Before playback, the operator started a calibration session and aligned virtual speaker markers to the physical loudspeakers using \emph{Spatial DJ}'s AR controller. The system recorded the resulting 3D layout in a room-aligned coordinate frame and persisted it in the database for reuse. The operator then assembled a scene by assigning stems to point-source or ambient audio objects, and routing them through the input channels of the AoIP interface to the \emph{Spatial DJ} backend. These deployments do not establish perceptual optimality or equivalent rendering quality across configurations---rather, they demonstrate the core practical objective of the system: applying a common workflow for on-site calibration and real-time control across heterogeneous loudspeaker geometries using the same representation and mapping rules.

\section{Discussion}
\emph{Spatial DJ} treats the loudspeaker layout as a varying system input rather than a fixed assumption. Instead of preparing multiple versions of a spatial mix for different loudspeaker configurations, creators first acquire the physical layout and then manipulate audio objects directly within the same room-aligned coordinate frame in which they are reproduced. Our approach responds to practical challenges previously identified in spatial audio production. Peters et al. \cite{Peters2011CurrentTechnologies} found that insufficient venue calibration time and the technical complexity of multichannel reproduction often lead creators to under-utilize the available spatial features. \emph{Spatial DJ} addresses this issue by incorporating layout acquisition directly into the spatial audio production process. Once the encountered configuration is registered, creators can render, audition, and manipulate spatial audio objects within that physical environment. Our four representative deployments demonstrate workflow portability across stereo, surround 7.1, periphonic 7.1.4, and irregular venue configurations. They do not establish perceptual equivalence across those systems as differences in loudspeaker coverage, room acoustics, and audience position may substantially alter the resulting experience, leaving the perceptual transmutation of a soundscape across heterogeneous venues an open question for future studies. 

Our technical evaluation shows that this workflow can recover loudspeaker geometry with centimeter-scale accuracy and propagate control updates to the renderer with low latency in the evaluated environment. The calibration errors showed lower variation along the vertical $y$-axis than the horizontal $x$- and $z$-axes. One possible explanation is that the vertical reference was relatively consistent across sessions, as each operator's standing height was unlikely to vary substantially between repeated calibrations. In contrast, small differences in the HMD's position and orientation at the AR session initialization may have introduced greater variation in the horizontal axes, since the initial headset pose contributes to the alignment of the room-aligned coordinate frame. It is worth noting that these deviations remained small relative to the physical dimensions of the loudspeakers used in the evaluation.

Our listener study showed that participants' descriptions of perceived motion did not map one-to-one onto the predefined trajectories used to emulate authored spatial motion: orbiting trajectories were frequently identified as elliptical, yet listeners varied in their exact descriptions, and even the diffuse baseline occasionally elicited apparent horizontal or vertical one-dimensional motion. This finding further motivates the co-location of authoring and on-site audition: creators may need to hear how a trajectory is realized through the actual loudspeaker geometry rather than relying solely on their expectations. \emph{Spatial DJ} enables this through real-time control, allowing creators to manipulate spatial audio objects directly in the venue and hear the resulting changes through the physical loudspeaker system with minimal delay. Our technical evaluation showed a median control latency of 8.07\,ms, below the 10\,ms latency guideline commonly used for responsive musical interaction
\cite{https://doi.org/10.5281/zenodo.1176382}. However, the authoring capability was not directly evaluated in our listener study, which used a controlled set of predefined trajectories for experimental perceptual consistency. This leaves a future research avenue for how creators might use \emph{Spatial DJ} when authoring and auditioning free-hand trajectories and other spatial rendering configurations.

Across the tested musical excerpts, configurations containing orbiting point-source audio objects received higher ratings of perceived immersion, energy, and novelty than the fully diffuse baseline. The coupling of position and spread parameters is inherent to the way the conditions were constructed: an ambient object is distributed equally across the loudspeaker array, so changing its position does not produce a trajectory. Introducing spatial motion therefore requires the object to be rendered as a localized point source, whose position can then vary over time. Within this design, the results show that spatial rendering configurations enabled by \emph{Spatial DJ} can produce substantial experiential differences, supporting trajectory as a useful authoring dimension \cite{Deacon2023SpatialDesign, dearvr2024}.

The results also indicate that spatial trajectories may be most useful when they can be assigned independently to different musical layers. Perhaps a bit counterintuitively, vocal focus ratings were highest when the instrumental audio object followed an orbit while the vocal layer remained diffuse, showing that the spatial rendering configuration applied to one musical layer can be associated with how another layer is perceived. This motivates authoring interfaces that let creators assign spatial rendering configurations independently to constituent stems and raises a broader perceptual question concerning how spatialization interacts across layers, inviting further research. At the same time, the dual orbit condition suggests that increasing spatial complexity does not automatically translate into an experiential advantage. Even though listeners more frequently reported perceiving more complex spatial movement in this condition, the reported comparisons did not establish a statistically significant advantage over single-layer motion for immersion, motion clarity, energy, or novelty. Together with comments describing dual motion as both memorable and potentially overwhelming during prolonged listening, our results suggest that authoring tools should make spatial complexity controllable. The short trials also leave open how these spatial rendering configurations behave over longer listening periods. We did not detect a statistically significant effect on distraction or discomfort during the 30-second excerpts, yet some open-ended comments left open the possibility of potential negative effects if motion is sustained for longer periods. Thus, the absence of significant rating difference should not be interpreted as evidence that complex spatial motion remains comfortable over longer performances. Temporal duration of spatial effects may be an important authoring parameter for a future study.

A natural next step is therefore to evaluate \emph{Spatial DJ} not only as a renderer but as an interactive authoring system, allowing users to create freehand trajectories directly through the AR interface. Rather than replaying predefined trajectories, future studies could ask creators to construct, refine, and audition their own spatial behaviors directly in the physical reproduction environment. Such studies could examine how creators use room-aligned interaction when the rendered result depends on the particular loudspeaker geometry, and how this affects authoring time, workload, perceived control, and the spatial behaviors they produce. This would also extend prior gestural authoring work in binaural XR \cite{audiomixr_2025_geleta} and fixed room-scale loudspeaker installations \cite{the_boomroom_2014} to a setting in which audio is reproduced through physical loudspeakers whose geometry is acquired on site and treated as a variable. A further extension would be to explore multi-user scenarios, where several creators or performers jointly manipulate spatial audio objects within the same physical loudspeaker environment.

\section{Conclusion}
We presented \emph{Spatial DJ}, a system for spatial audio rendering and control system that treats the physical loudspeaker layout as a varying system input rather than a fixed assumption. The system integrates on-site layout acquisition in which operators visually align virtual speaker markers to physical loudspeakers using AR HMD tracking, and the system records and persists the recovered 3D geometry in a room-aligned coordinate frame. Given this geometry, the backend maps lightweight object states to per-speaker gains weighted by distance and constant power normalization. In a controlled 18-loudspeaker environment, we quantified calibration accuracy and repeatability of the AR controller, and measured control latency across the evaluated scene complexities. An 18-participant listener study further showed systematic perceptual differences among spatial rendering configurations applied independently to vocal and instrumental layers, while highlighting that realized spatial trajectories do not always map directly onto perceived motion. Finally, representative deployments across stereo, surround 7.1, periphonic 7.1.4, and an asymmetric eight-speaker nightclub configurations demonstrated reuse of the same workflow across heterogeneous layouts. Altogether, these results support \emph{Spatial DJ} as a step toward spatial audio authoring workflows in which creators can build and audition their spatial soundscapes directly within the physical environments in which they will be reproduced.

\subsubsection*{Acknowledgments}
We are grateful to Craig Linssen and the Apple Developer Team for their continued support and hardware sponsorship. We thank Erich Krey and Yvan Grabit from Steinberg for the insights and pointing us toward practical OSC workflows. We also thank Andy Wilson who kindly offered us a laboratory space at Microsoft Research to test the early stages of our prototype. We want to extend our gratitude to Artur Geleta who helped us with audio equipment and testing venues.

\subsubsection*{Ethics and Privacy Statement}
In accordance with guidance from the Spanish Research Ethics Committee (Comité Español de Ética de la Investigación, CEEI), 
the study protocol was submitted for independent ethical review and was determined to be exempt by Universidad Miguel Hernández de Elche (UMH). 
The study was conducted in compliance with the European General Data Protection Regulation (GDPR). 
All participants provided informed consent prior to participation and were informed of the study's purpose, procedures, and their right to withdraw at any time without penalty. No directly identifying information was collected.

\bibliographystyle{abbrv}
\bibliography{sample-base}

\newpage
\appendix

\section{Listener Study Questionnaire} \label{appendix}

\subsection{Consent and Demographics}

\begin{enumerate}
    \item \textbf{Would you like to participate in this study as described above?}
    
    \begin{itemize}[label=--]
        \item Yes, I would like to participate.
        \item No. (End of participation)
    \end{itemize}

    \item \textbf{Gender} (Optional)
    
    \begin{itemize}[label=--]
        \item Female
        \item Male
        \item Other
    \end{itemize}

    \item \textbf{Age range} (Optional)
    
    \begin{itemize}[label=--]
        \item 18--24
        \item 25--34
        \item 35--44
        \item 45--54
        \item 55+
    \end{itemize}

    \item \textbf{Do you have any hearing difficulties?}
    
    \begin{itemize}[label=--]
        \item No difficulty hearing
        \item Some difficulty hearing
        \item Many difficulties with hearing
        \item I cannot hear at all
        
    \end{itemize}
\end{enumerate}

\subsection{Trial Questionnaire}

The following questions were presented immediately after each listening trial. For Questions 1--6, participants responded using a seven-point Likert scale:

\begin{center}
\begin{tabular}{cl}
1 & Strongly disagree \\
2 & Disagree \\
3 & Somewhat disagree \\
4 & Neutral \\
5 & Somewhat agree \\
6 & Agree \\
7 & Strongly agree \\
\end{tabular}
\end{center}

\begin{enumerate}
    \item \textbf{Immersion}\\
    \enquote{I felt surrounded or immersed in the sound.}

    \item \textbf{Motion clarity}\\
    \enquote{The spatial movement of the sound was clear and easy to follow.}

    \item \textbf{Vocal focus}\\
    \enquote{It was easy to focus on the vocals (voice).}

    \item \textbf{Energy / intensity}\\
    \enquote{This audio effect made the soundscape feel more energetic or intense.}

    \item \textbf{Novelty}\\
    \enquote{This audio effect felt novel or unusual.}

    \item \textbf{Distraction / discomfort}\\
    \enquote{This audio effect was distracting or unpleasant.}

    \item \textbf{Perceived motion}\\
    \enquote{Which option best describes your perception? Select one or more.}

    \begin{itemize}
        \item Diffuse or dispersed
        \item Elliptical motion
        \item Horizontal motion
        \item Vertical motion
        \item Two moving layers / objects
        \item Not sure
        \item Other: \rule{3cm}{0.4pt}
    \end{itemize}
\end{enumerate}

\subsection{Post-Study Questions}

After completing all listening trials, participants responded to the following open-ended questions:

\begin{enumerate}
    \item \enquote{Did your perception of the same excerpt change when you heard it with a different audio effect?}

    \item \enquote{Considering all the excerpts, which differences in spatial perception were most noticeable or memorable to you?}

    \item \enquote{Was there anything important about your experience that was not captured by the questionnaire?}
\end{enumerate}

\end{document}